\documentstyle[preprint,aps]{revtex}  % PREPRINT
\begin{document}
\draft         % PRINTS PACS NUMBER
\preprint{IFUSP-P 1083}

\title{ Signals for Vector Leptoquarks in Hadronic Collisions}

\author{ J.\ E.\ Cieza Montalvo $^*$ and O.\ J.\ P.\ \'Eboli $^\dagger$}

\address{ Instituto de F\'{\i}sica, Universidade de S\~ao Paulo \\
C.P. 20516, 01452-990 S\~ao Paulo, Brazil}

\date{\today}

\maketitle

\begin{abstract}
We analyze systematically the signatures of vector leptoquarks in
hadronic collisions. We examine their single and pair productions, as
well as their effects on the production of lepton pairs.  Our results
indicate that a machine like the CERN Large Hadron Collider (LHC) will
be able to unravel the existence of vector leptoquarks with masses up
to the range of $2$--$3$ TeV.
\vskip 2.5cm
\begin{center}
{\em Submitted to Phys.\ Rev.\ D}
\end{center}
\end{abstract}
%\pacs{}

\newpage

%^^^^^^^^^^^^^^^^^^^^^^^^^^^^^^^^^^^^^^^^^^^^^^^^^^^^^^^^^^^^^^^^^^^^^^^

\section{Introduction}

Many extensions of the Standard Model predict the existence of
leptoquarks, which are color triplet particles carrying simultaneously
leptonic and baryonic number and can have spin $0$ or $1$.  For
instance, some composite models postulate a preonic sub-structure
where quarks and leptons have some common constituents \cite{comp,af}.
These models exhibit a very rich spectrum which includes excited
states of the known particles, as well as new particles possessing
unusual quantum numbers, such as leptoquarks and dileptons.
Leptoquarks are also present in extensions of the standard model that
treat quarks and leptons in the same basis and, consequently, allow
the existence of new particles mediating quark-lepton transitions.
This class of models includes grand unified theories \cite{gut},
technicolor models \cite{tec}, and superstring-inspired models
\cite{e6}.

The production and signals of scalar leptoquarks have already been
analyzed in the literature for $ep$ \cite{wud,v:ep}, hadronic
\cite{lepto:ha}, $e^+e^-$ \cite{lepto:ee}, and $e\gamma$ \cite{egamma}
colliders. However, previous studies of vector leptoquarks ($V$)
considered only $e^+e^-$ \cite{v:egm,v:egb} and $ep$ colliders
\cite{v:ep}.  In this work, we study the capability of hadron
colliders to unravel the existence of vector leptoquarks. We analyze
their pair production via gluon--gluon and quark--antiquark fusion ($g
(q)~+ g (\bar{q})~ \rightarrow V + \bar{V} $), single production in
association with a lepton ($g + q \rightarrow V + \ell \text{, }\ell =
e^\pm, \mu^\pm, \nu$), and their effects in the production of lepton pairs ($ q
+ \bar{q} \rightarrow \ell^+ + \ell^-$). Our results show that a
machine like the LHC can expand considerably the range of leptoquark
masses and couplings that can be explored.

It is interesting to notice that pair production of vector leptoquarks
is essentially model independent since it is related to the strong
interactions.  On the other hand, single production and indirect
effects are to a great extent model dependent because they are due to
unknown interactions.  Moreover, these three signals for vector
leptoquarks are complementary since they allow us not only to reveal
their existence but also to determine their
properties such as mass, spin, and couplings to quarks and leptons.

There have been several direct searches for leptoquarks in
accelerators.  Analyzing the decay of the $Z$ into a pair of on-shell
leptoquarks, the LEP experiments established a lower bound $ M_{lq}
\gtrsim 44$ GeV for scalar leptoquarks \cite{LEP,DELPHI}, and a
similar bound should hold for vector ones. This limit can be improved
by looking for $Z$ decays into an on-shell and an off-shell
leptoquarks, resulting in $M_{lq} \gtrsim 65-73$ GeV \cite{DELPHI}.
The search for scalar leptoquarks decaying into an electron-jet pair
in $p\bar{p}$ colliders constrained their masses to be $M_{lq} \gtrsim
113$ GeV \cite{PP}. Furthermore, the experiments at HERA \cite{HERA}
placed limits on their masses and couplings, leading to $M_{lq}
\gtrsim 92-184$ GeV depending on the leptoquark type and couplings.

Leptoquarks are also tightly constrained by low energy experiments
\cite{cons:1,cons:2}, such as the proton lifetime measurements and the
study of rare decays associated to flavor changing neutral currents.
In order to avoid the strong low energy bounds we assumed that the
leptoquark interactions conserve lepton and baryon number in addition
to electric charge and color, and that their couplings are chiral
\cite{cons:1}. Furthermore, leptoquarks that couple to more then one
fermion generation are constrained to be very heavy, therefore,
we shall study vector leptoquarks that couple only to one
family.

For the sake of definiteness, we shall consider the vector leptoquarks
inspired by the Abbott--Farhi model \cite{af}, which is a confining
version of the Standard Model and is described by a Lagrangian which
has the same general structure of the Standard Model one.
Nevertheless, the choice of parameters is such that no spontaneous
symmetry breaking occurs and the $SU(2)_L$ gauge interaction is
confining.  In this model, the physical left-handed fermions are bound
states of a preonic scalar and a preonic left-handed fermion, while
the vector bosons are P--wave bound states of the scalar preons. The
vector leptoquarks predicted by this model are color triplets with
electric charge $Q_V=-2/3$ and weak isospin $0$.

This model cannot be analyzed perturbatively since it is strongly
interacting in the energy scale of interest. Therefore, we shall
parametrize the interaction between vector leptoquarks ($V_\mu^{ab}$)
and physical left-handed fermions by an effective Lagrangian
\cite{wud} given by
\begin{equation}
{\cal L}_{\rm int} = -F{e\over 2\sqrt{2}\sin \theta_W}
\left( {V_\mu^{ab}}^\dagger
\bar L^{a} \gamma^\mu  L^b + \mbox{h.c.} \right ) \; ,
\label{lint}
\end{equation}
where $L^a$ are physical left-handed doublets, $a$ ($b$) is a flavor
index, $\theta_W$ is the weak mixing angle, and the parameter $F$ is a
measure of the strength of this interaction compared to the $Wq\bar
q^\prime$ vertex.  The above Lagrangian is a prototype of a wide class
of models \cite{v:ep} presenting the interaction of color triplet,
charged vectors with $B$ and $L$ numbers.  According to (\ref{lint}),
vector leptoquarks couple to both upper or lower components of the
lepton and quark doublets with the same strength.

We can estimate the size of the coupling $F$ imposing that unitarity is
respected at tree level \cite{corn}. For instance, the process $e^+
e^- \rightarrow V \bar{V}$ violates unitarity at high energies for
arbitrary values of the couplings.  However, this can be avoided by
choosing $F= \sqrt{|Q_V|}=\sqrt{\frac{2}{3}}$ \cite{v:egm}.

Since the vector leptoquarks are color triplets, it is natural to
assume that they interact with gluons.  To obtain their couplings to
gluons we substituted $\partial_\mu$ by the covariant derivative
$D_\mu = \partial_\mu + i g_s \lambda^a_{jk} G^a_\mu /2$ in the vector
leptoquark kinetic lagrangian and introduced an anomalous color
magnetic moment $\kappa$. For $\kappa\ne 1$, unitarity is not
respected at tree level by this model. Therefore, we must either
impose $\kappa=1$ or introduce a new cutoff mass scale $\Lambda$.  We
assumed that $\kappa=1$, since this leads to conservative estimates
for the cross sections.

The main decay modes of vector leptoquarks are into pairs $\ell q$ and
$\nu q^\prime$, thus, their signal is a charged lepton plus a
jet, or a jet plus missing energy. Using the couplings given above we
obtain that the width of a vector leptoquark is
\begin{equation}
\Gamma_V = \frac{\alpha F^2}{4\sin^2\theta_W} M_V \; ,
\end{equation}
where we neglected all fermion masses and summed over all possible
decay channels.

The outline of this paper is the following.  Section II contains the
study of the production of vector leptoquark pairs via
quark--antiquark and gluon--gluon fusion, which leads to conservative
estimates of the discover potential of vector leptoquarks in the next
generation of hadronic colliders.  In Sec.\ III we analyze the single
production of vector leptoquarks through quark--gluon fusion,
exhibiting our results for first and second generation vector
leptoquarks.  The analysis of the indirect signals for leptoquarks is
contained in Sec.\ IV: One way to look for vector leptoquarks is
through their effects on the production of lepton pairs at high
energies, since they can be exchanged in the $t$-channel.  Our
conclusions are drawn in Sec.\ V.

%^^^^^^^^^^^^^^^^^^^^^^^^^^^^^^^^^^^^^^^^^^^^^^^^^^^^^^^^^^^^^^^^^^^^^^^^^^^^^

\section{Pair Production of Vector Leptoquarks}

Pair production of new particles is, to a good approximation, a model
independent process since it usually proceeds through well known
strong and electroweak interactions. Consequently, this process leads
to more reliable estimates of the possibility of discovering new
physics. In a hadronic collider there are two main mechanisms for the
production of vector leptoquark pairs: gluon--gluon fusion ($g + g
\rightarrow V + \bar{V}$) and quark--antiquark fusion ($q + \bar{q}
\rightarrow V + \bar{V} $).

In lowest order, the Feynman diagrams contributing to the production
of vector leptoquark pairs via gluon--gluon fusion are the exchange of
gluons in the $s$-channel, vector leptoquarks in the $t$- and
$u$-channels, and a contact interaction ($ggVV$), being its cross
section for $\kappa=1$ given by \cite{boris}
\begin{eqnarray}
\frac{d\hat{\sigma}_{gg}}{d\cos\theta} =&&
\frac{\pi\alpha_s^2}{16} \frac{\beta}{\hat{s}} \left ( \frac{4}{3} -
3 \frac{t_M u_M}{\hat{s}^2} \right ) \times
\label{xsec:gg}
\\
&& \left [ 3 + 2 \frac{\hat{s}^2}{t_M u_M}
\left ( \frac{\hat{s}^2}{t_M u_M} - 2 \right ) + 6
\frac{\hat{s} M_V^2}{t_M u_M}
\left ( \frac{\hat{s} M_V^2}{t_M u_M}  - 1 \right ) \right ] \; ,
\nonumber
\end{eqnarray}
where $\sqrt{\hat{s}}$ is the invariant mass of the subprocess, $t_M =
- \hat{s} (1 - \beta \cos\theta) /2$, $u_M = - \hat{s} (1 + \beta
\cos\theta) /2$, $\theta$ is the angle between the vector leptoquark
and the incident gluon in the subprocess center of mass frame, and
$\beta = \sqrt{1 - 4 M_V^2 /\hat{s}}$ is the leptoquark velocity in
this frame. This cross section has a good high energy behavior and,
for large values of $\hat{s}$, it approaches a constant value
proportional to $1/ M_V^2$ due to the exchange of a vector
particle in the $t$-channel.

The quark--antiquark fusion production of vector leptoquark pairs
takes place by the exchange of a gluon in the $s$-channel and a lepton
in the $t$-channel. However, this last contribution is expected to be
much smaller than the gluon exchange since the leptoquark couplings to
quarks and leptons are expected to have a weak interaction strength --
that is, it is natural to have $F$ of the order of the unit.
Therefore, we restrict ourselves to the $s$-channel production only
\cite{he}, and the elementary cross section for this process is
\cite{boris}
\begin{eqnarray}
\frac{ d\hat{\sigma}_{q\bar{q}} }{ d \cos \theta }  &&=
\frac{9 \pi \alpha_s^2}{2} \frac{\beta}{\hat{s}} \times \\
&& \left \{\frac{1}{4} \left ( 1 + \beta^2 \right )^2
\frac{\hat{s}}{M_V^2} -1+\beta^2 +\frac{1}{4} \left ( \frac{t_M
u_M}{\hat{s}^2} - \frac{M_V^2}{\hat{s}}
\right ) \left [ 8 + \frac{1}{4} \left ( 1 + \beta^2 \right )^2
\frac{\hat{s}^2}{M_V^4} \right ] \right \} \; . \nonumber
\end{eqnarray}

In order to obtain the inclusive total cross section for the
production of pairs of vector leptoquarks ($pp \rightarrow V \bar{V}
X$) we must fold the proton structure functions with the elementary
cross sections above
\begin{eqnarray}
\sigma_{\text{pair}} = \int dx_1 dx_2~ \Biggl [ &&
g(x_1, Q^2) g(x_2,Q^2) \hat{\sigma}_{gg} (\hat{s}) \nonumber \\
&& +  \left ( q(x_1, Q^2) \bar{q}(x_2, Q^2) + q(x_2, Q^2) \bar{q}(x_1, Q^2)
\right ) \hat{\sigma}_{q\bar{q}} (\hat{s}) \Biggr ] \; ,
\end{eqnarray}
where $g$ and $q$ are the relevant gluon and quark structure
functions, and $\hat{s} = x_1 x_2 s$ with $\sqrt{s}$ being the machine
center of mass energy.  Figure \ref{vvpair} exhibits our results for
the total cross section for the production of vector leptoquark pairs
as a function of its mass at the LHC (taking $\sqrt{s} = 15.4$ TeV),
where we imposed a rapidity cut $|y_{V(\bar{V})}| < 2.5$ and used the
set I of structure functions given in Ref.\ \cite{do} evaluated at the
scale $Q^2=\hat{s}$. As expected, the gluon--gluon--fusion
contribution dominates over the quark--antiquark--fusion one at such
high energies due to the large gluon-gluon luminosity. Notice that the
above results hold true for any kind of vector leptoquark since we
used only their strong interaction properties in the calculations.

The signal for leptoquark pair production is either $jj\ell^+\ell^-$
($j=$ jet), $jj\ell^\pm{\rm p} \hspace{-0.53 em} \raisebox{-0.27
ex}{/}_T $, or $jj{\rm p} \hspace{-0.53 em} \raisebox{-0.27 ex}
{/}_T$, where $\ell^\pm = e^\pm$ or $\mu^\pm$.  Certainly the last
signal will be emersed in a QCD background and will be very hard to
extract. The other two, mainly the first, have a very good chance to
be observed since they have the striking feature of producing charged
lepton--jet pairs with a given invariant mass, however, a careful
studied of the QCD and electroweak backgrounds is needed to know
whether this kind of events can be seen or not \cite{we}.
Nevertheless, in order to have a rough estimate of the potential of
the LHC to unravel the existence of vector leptoquarks via their pair
production, we required the production of 100 of such pairs per year.
Taking into account the expected integrated luminosity for the LHC
($10^5$ pb$^{-1}$/yr), we obtain that this machine will be able to
discover first and second generation vector leptoquarks with masses up
to $2.1$ TeV.

%^^^^^^^^^^^^^^^^^^^^^^^^^^^^^^^^^^^^^^^^^^^^^^^^^^^^^^^^^^^^^^^^^^^^^^^

\section{Single Production of Vector Leptoquarks}

The production of a single leptoquark in a hadronic collider must be
associated with a lepton due to lepton number conservation. The
subprocess $ q + g \rightarrow V + \ell$ proceeds through the exchange
of a leptoquark in the $t$-channel and a quark in the $s$-channel.
This is a model dependent reaction since it involves unknown
leptoquark--quark--lepton interactions. Using the model described in
the introduction, it is easy to obtain that the elementary cross
section for this process is
\begin{eqnarray}
&& \frac{d\hat{\sigma}_{\text{single}}}{d\hat{t}}  =
\frac{\pi \alpha \alpha_s F^2}{12 \sin^2\theta_W M_V^2}~
\frac{1}{\hat{s}^2 (\hat{s}+\hat{t})^2} \times \\
&&  \left \{ 2 \beta^2 \hat{s}^3 \left ( 3 - 2 \beta \right )
+ \hat{s}^2 \hat{t}
\left ( -6 + 15 \beta - 2 \beta^2 - 2 \beta^3  \right )
+ \hat{s} \hat{t}^2
\left ( -1 + 8 \beta - 2 \beta^2 \right )
+ \hat{t}^3 \left ( 1 + \beta \right ) \right \} \; ,
\nonumber
\end{eqnarray}
where $\hat{t} = M_V^2 - (\hat{s}+M_V^2) (1-\beta \cos \theta ) /2$,
$\theta$ is the angle between the vector leptoquark and the incident
quark in the subprocess center of mass frame, and $\beta = 1 -
M_V^2/\hat{s}$ is the leptoquark velocity in this frame.

In order to obtain the total cross section for the inclusive process
$pp \rightarrow V \ell  X$ we must evaluate
\begin{equation}
\sigma_{\text{single}} = \int dx_1 dx_2~ \left [ g(x_1;Q^2)
q(x_2, Q^2) + g(x_2;Q^2) q(x_1;Q^2) \right ] \hat{\sigma}_{\text{single}}
(x_1x_2s) \; ,
\end{equation}
where $g$ is the gluon structure function and $q$ is the density of
down (up) quarks for $\ell=e^\pm$ ($\ell=\nu_e$) assuming a first
generation leptoquark. In the case of a second generation leptoquark,
$q$ is the strange (charm) structure function for $\ell=\mu^\pm$
($\ell=\nu_\mu$). Figure \ref{vlpair} summarize our results for the
total cross section for the production of a vector leptoquark in
association with charged leptons and neutrinos, where
we imposed a rapidity cut on the charged particles $|y_{V(\ell^\pm)}|
< 2.5$ and considered first and second generation vector leptoquarks.
As expected, the cross section for producing pairs $V\nu_e$ is
approximately a factor of $2$ larger than the one for the production
of $Ve^\pm$ since they are associated to the up and down structure
functions respectively. Since the strange content of the proton is
larger than the charm one, the cross section for production of pairs
$V\mu^\pm$ is larger than the one for pairs $V\nu_\mu$.

Taking into account that the decay of the leptoquark can give rise to
a pair electron-jet or a jet plus missing energy, three different
signals are possible for its single production: $j\ell^+\ell^-$,
$j\ell^\pm {\rm p} \hspace{-0.53 em} \raisebox{-0.27 ex}{/}_T$, and $j
{\rm p} \hspace{-0.53 em} \raisebox{-0.27 ex}{/}_T$, where
$\ell^\pm=e^\pm$ or $\mu^\pm$.  This last signal has an incisive
feature since it is a monojet event. However, it is going to be buried
in the background produced by standard-model processes, like the
production of $Z~(\rightarrow \nu\bar{\nu})$ associated with a jet,
which are expected to have a much higher cross sections \cite{ehlq}.
On the other hand, the other two possible signals are very striking
since they exhibit pairs $\ell^\pm j$ with (approximately) the same
invariant mass. {\em A priori\/}, it is hard to estimate the
background for these signals since it is due to the QCD production of
heavy quarks, or $W$ and $Z$ in association with jets, thus requiring
further Monte Carlo work to determine their size \cite{we}.

In order to have an approximate idea of the potential of the LHC to
discover vector leptoquarks via their single production, we required
100 events per year exhibiting a leptoquark decaying into a
$\ell^\pm j$ pair.  Using the expected luminosity for the LHC and that
the branching ratio of vector leptoquarks to $\ell^\pm j$ pairs is
$1/2$, we obtain that the LHC will able to probe first (second)
generation vector leptoquark masses up to $2.7$, $3.0$, and $3.5$
($1.9$, $2.1$, and $2.4$) for couplings $F=0.5$, $\sqrt{2/3}$, and
$1.5$ respectively.

%^^^^^^^^^^^^^^^^^^^^^^^^^^^^^^^^^^^^^^^^^^^^^^^^^^^^^^^^^^^^^^^^^^^^^^^^^^

\section{Indirect Signals for Vector Leptoquarks}

The existence of leptoquarks can also be investigated through their
effects as an intermediate state in the inclusive dilepton production
($pp \rightarrow \ell^+ \ell^- X$), where they lead to a new
$t$-channel contribution in addition to the usual exchange of $\gamma$
and $Z$ in the $s$-channel. The color averaged cross section for the
subprocess $q \bar{q} \rightarrow \ell^+ \ell^-$ is, in the center of
mass of the subprocess,
\begin{eqnarray}
\frac{d\hat\sigma}{d\Omega}(\hat{s}) = && \frac{\alpha^2}{12\hat{s}}
\Biggl \{ Q^2 (1 + \cos^2\theta)
%\nonumber \\ &&
+ \frac{1}{16\sin^4\theta_W\cos^4\theta_W} ~
\frac{\hat{s}^2}{(\hat{s}-M_Z^2)^2+\Gamma_Z^2 M_Z^2} \nonumber \\
&& \times \left
[ \left ({C_V^e}^2 + {C_A^e}^2 \right ) \left ( {C_V^q}^2 + {C_A^q}^2
\right ) (1+\cos^2\theta) +8 C_V^e C_A^e C_V^q C_A^q \cos\theta \right ]
\nonumber \\
&& - \frac{Q}{2 \sin^2\theta_W \cos^2\theta_W }~\frac{\hat{s}(\hat{s}-M_Z^2)}
{(\hat{s}-M_Z^2)^2+\Gamma_Z^2 M_Z^2} \left [
C_V^eC_V^q (1+\cos^2\theta) + 2 C_A^eC_A^q \cos\theta \right ]
\nonumber \\
&&+ \frac{F^2}{\sin^2\theta_W}~ \frac{(1+\cos\theta)^2}{\cos\theta-\eta}
\Biggl [ \frac{F^2}{4\sin^2\theta_W}~\frac{1}{\cos\theta-\eta}
+ \frac{Q}{2} \\
&&- \frac{1}{8\sin^2\theta_W\cos^2\theta_W}~ (C_V^q+C_A^q)(C_V^e+C_A^e)
{}~\frac{\hat{s}(\hat{s}-M_Z^2)}
{(\hat{s}-M_Z^2)^2+\Gamma_Z^2 M_Z^2} \Biggr ] \Biggr \} \; ,
\nonumber
\end{eqnarray}
where $M_Z$ is the mass of the $Z$ boson and $\eta = 1 +2M_V^2/s$.
According to our conventions the charge of a quark is $Qe$ ($e> 0$), $
C_V=I_z-2Q\sin^2\theta_W$, and $C_A = I_z$.

The exchange of a vector particle in the $t$-channel modifies the high
energy behaviour of this subprocess: within the scope of the standard
model this elementary cross section decreases as the center of mass
energy increases, however, the new contribution alters this behaviour,
yielding a constant cross section at high energies
\begin{equation}
\sigma_{\rm limit} (q \bar{q} \rightarrow \ell^+ \ell^-)
\simeq \frac{\pi \alpha \alpha_s F^4}{12 \sin^4\theta_W M_V^2} \; .
\end{equation}

The effect of this modification to the process $pp \rightarrow
e^+e^-X$ due to a first generation vector leptoquark can be seen on
Fig.\ \ref{llpair:m}, which displays the invariant mass distribution
of dilepton pairs at the LHC for a vector leptoquark of mass $M_V = 1$
TeV and several values of the coupling $F$. From this figure we can
see that the presence of the leptoquark leads to a dramatic increase
of the production of dilepton pairs at high invariant masses.

In order to estimate the capability of the LHC to search for vector
leptoquarks, we evaluated the total cross section for the production
of dileptons with an invariant mass greater than $300$ GeV, exhibiting
our results for first and second generation vector leptoquarks in
Figs.\ \ref{llpair:tot} and \ref{llpair2:tot} respectively. Demanding
that the anomalous production of high invariant mass dileptons have a
statistical significance of $5\sigma$, we obtain that the largest
first (second) generation vector leptoquark mass accessible at the LHC
is $1.3$, $2.1$, and $4.0$ ($0.8$, $1.4$, and $2.6$) TeV assuming that
$F=0.5$, $\sqrt{2/3}$, and $1.5$ respectively.

%^^^^^^^^^^^^^^^^^^^^^^^^^^^^^^^^^^^^^^^^^^^^^^^^^^^^^^^^^^^^^^^^^^^^^^^^^^

\section{Conclusions}

The discover of leptoquarks is without any doubt a striking signal for
the existence of life beyond the standard model. In this work we
analyzed the potential of hadronic colliders, like the LHC, to unravel
the existence of vector leptoquarks. Our results for the discovery
limits at the LHC are summarized in Table \ref{tab:sum}, from where we
see that such a machine can extend considerably the range of
leptoquark masses and couplings that can be probed.

It is interesting to notice that the three signals for vector
leptoquarks that we studied are complementary, {\em i.e.\/} their
analysis allow us to shed some light on the leptoquark type and its
couplings to quarks and leptons. For instance, the pair production,
since is basically model independent, can be used to establish their
existence. On the other hand, the single production rates, together
with the knowledge of the charged lepton observed from their decays,
can be used to determine to which quarks the leptoquarks couple.
Moreover, the production of high invariant mass lepton pairs can
further confirm the signal and help to pinpoint the value of the
couplings between leptoquarks, quarks and leptons.

\vskip 18pt

\noindent {\em Note added:} During the time that we were
writing this work we became aware of Ref.\ \cite{v:had}
that analyzes the vector leptoquark pair production in
hadronic colliders. We verified that our results for this
process agree.

\acknowledgments

One of us (O.J.P.E.) would like to thank the kind hospitality of the
Institute for Elementary Particle Research, University of
Wisconsin--Madison, where the final part of this work was done.  We
are also grateful to M.\ C.\ Gonz\'alez-Garc\'{\i}a for a critical
reading of the manuscript. This work was partially supported by
Conselho Nacional de Desenvolvimento Cient\'\i fico e Tecnol\'ogico
(CNPq), and Funda\c{c}\~ao de Amparo \`a Pesquisa do Estado de S\~ao
Paulo (FAPESP).

%^^^^^^^^^^^^^^^^^^^^^^^^^^^^^^^^^^^^^^^^^^^^^^^^^^^^^^^^^^^^^^^^^^^^^

%^^^^^^^^^^^^^^^^^^^^^^^^^^^^^^^^^^^^^^^^^^^^^^^^^^^^^^^^^^^^^^^^^^^^^^^^

%
%       Table
%

\begin{table}
\begin{displaymath}
\begin{array}{||c|c|c|c||}
\hline
\hline
F   & V \bar{V} & V\ell & \ell^+\ell^- \\
\hline
0.5 & 2.1/2.1 & 2.7/1.9 & 1.3/0.8 \\
\protect\sqrt{\frac{2}{3}} & 2.1/2.1 & 3.0/2.1 & 2.1/1.4 \\
1.5 & 2.1/2.1 & 3.5/2.4 & 4.0/2.6 \\
\hline
\hline
\end{array}
\end{displaymath}
\caption{Maximum first/second generation vector leptoquark masses that can be
 probed at the LHC for pair ($V\bar{V}$), single ($V\ell$), and
dilepton ($\ell^+\ell^-$) production, assuming an integrated
luminosity of $10^5$ pb$^{-1}$ /yr. }
\label{tab:sum}
\end{table}

%^^^^^^^^^^^^^^^^^^^^^^^^^^^^^^^^^^^^^^^^^^^^^^^^^^^^^^^^^^^^^^^^^^^^^^^^

%
% Figure Caption
%

\begin{figure}
\caption{Cross section for the production of vector leptoquarks pairs
as a function of $M_V$ at the LHC: the dashed, dotted, and
solid lines stand for $q\bar{q}$, $gg$ and total contributions. We
imposed the rapidity cut $|y_V|<2.5$. }
\label{vvpair}
\end{figure}

\begin{figure}
\caption{ Total cross section for the associated production of vector
leptoquarks and leptons ($e^\pm$, $\mu^\pm$, $\nu_e$, and $\nu_\mu$) as
a function of $M_V$ at the LHC.  The dashed and solid (dotted and
dash-dotted) stand for the production of a first (second) generation
leptoquark in association with $e^\pm$ and $\nu_e$ ($\mu^\pm$ and
$\nu_\mu$) respectively. We imposed the rapidity cut
$|y_{V{\ell^\pm}}| < 2.5$.}
\label{vlpair}
\end{figure}

\begin{figure}
\caption{ Invariant mass distribution for $e^+e^-$ pairs at the LHC: the solid,
dotted, dashed, and dot-dashed lines stand for $F=0$ (standard model),
$0.5$, $\protect\sqrt{2/3}$, and $1.5$ respectively. We assumed a
first generation leptoquark of mass $M_V = 1$ TeV and imposed that
$|y_{e^\pm}| < 2.5$.}
\label{llpair:m}
\end{figure}

\begin{figure}
\caption{ Total cross section for the production of
$e^+e^-$ pairs as a function of $M_V$ at the LHC, satisfying
$M_{e^+e^-} > 300$ GeV and $|y_{e^\pm}|< 2.5$. Conventions are the
same as in Fig.\ \protect\ref{llpair:m} }
\label{llpair:tot}
\end{figure}

\begin{figure}
\caption{ Same as in Fig.\ \protect\ref{llpair:tot} but
for $\mu^+\mu^-$ pairs. }
\label{llpair2:tot}
\end{figure}


\begin{references}
\bibitem{he} Taking into account only the $s$-channel gluon exchange
spoils the high energy behaviour of this cross section. The cure to
this problem requires the introduction of further interactions since
all kind of quarks participate in this reaction. We shall neglect this
fact, assuming that these new contributions are important only at very
high energy where the structure function of the proton is negligible.
\bibitem{careful} In this estimate we considered only the irreducible
background coming from the exchange of a $\gamma$ or a $Z$ in the
$s$-channel.  However, there are further backgrounds such as the
production of heavy quark pair which should be included in a more
accurate analysis.
\end{references}
\end{document}